# LLM Chatbot-Creation Approaches

Hemil Mehta
Department of Computer Science
North Carolina State University
Raleigh, NC, USA
hmehta@ncsu.edu

Tanvi Raut
Department of Computer Science
North Carolina State University
Raleigh, NC, USA
trraut@ncsu.edu

Kohav Yadav
Department of Computer Science
North Carolina State University
Raleigh, NC, USA
kyadav4@ncsu.edu

Edward F. Gehringer
Department of Computer Science
North Carolina State University
Raleigh, NC, USA
efg@ncsu.edu

*Abstract*—This full research-to-practice paper explores approaches for developing course chatbots by comparing low-code platforms and custom-coded solutions in educational contexts. With the rise of Large Language Models (LLMs) like GPT-4 and LLaMA, LLM-based chatbots are being integrated into teaching workflows to automate tasks, provide assistance, and offer scalable support. However, selecting the optimal development strategy requires balancing ease of use, customization, data privacy, and scalability. This study compares two development approaches: low-code platforms like AnythingLLM and Botpress, with custom-coded solutions using LangChain, FAISS, and FastAPI. The research uses Prompt engineering, Retrieval-augmented generation (RAG), and personalization to evaluate chatbot prototypes across technical performance, scalability, and user experience. Findings indicate that while low-code platforms enable rapid prototyping, they face limitations in customization and scaling, while custom-coded systems offer more control but require significant technical expertise. Both approaches successfully implement key research principles such as adaptive feedback loops and conversational continuity. The study provides a framework for selecting the appropriate development strategy based on institutional goals and resources. Future work will focus on hybrid solutions that combine low-code accessibility with modular customization and incorporate multimodal input for intelligent tutoring systems.

*Keywords—Artificial Intelligence, Learning Technologies, Intelligent Tutoring Systems, Chatbot Development*

## I. Introduction

### A. Background

Recent breakthroughs in Large Language Models (LLMs) such as GPT-4, Claude, LLaMA, and DeepSeek have significantly expanded the capabilities of conversational AI [1], [2]. These models support high-quality multi-turn dialogue, contextual memory, and few-shot learning enabling developers to construct intelligent chatbots without the need for training models from scratch [1]. The availability of model APIs, embedding techniques, and orchestration frameworks has also lowered the barrier to entry for creating domain-specific assistants [3], [4]. As a result, a diverse set of design strategies have emerged, including Prompt engineering, Retrieval-augmented generation (RAG), fine-tuning on curated datasets, and hybrid models that blend rule-based logic with generative outputs [4]. This shift in capability and accessibility has opened new frontiers for scalable, intelligent, and adaptable chatbot systems.

### B. Importance

In educational environments, LLM-powered chatbots offer transformative potential for both learners and instructors. These systems can automate instructional support by answering course-specific questions, assisting with assignments, and even summarizing student feedback during course evaluations [5], [6]. Their ability to function as always-available teaching assistants is especially beneficial in large or asynchronous courses, where providing individualized attention at scale is often impractical. By offloading routine inquiries and enabling personalized student interactions, LLM-based chatbots can enhance student engagement, reduce instructor workload, and improve access to learning resources. However, their development requires navigating complex trade-offs across usability, accuracy, implementation overhead, and deployment constraints such as cloud hosting, API rate limits, and support for large unstructured inputs like textbooks and code files [6]. While many studies discuss the technical capabilities of LLMs in educational chatbots, there is a gap in integrating these technologies with established pedagogical frameworks. Our work is grounded in the principles of scaffolded learning, formative feedback, and cognitive load reduction, all of which are essential for effective teaching assistant (TA) support. By embedding LLMs into course workflows, this research aims to operationalize these pedagogical goals in scalable and accessible ways particularly in large or online classes where personalized TA interaction is often limited.

### C. Purpose and Scope

This research investigates the design, architecture, and deployment strategies involved in developing LLM-powered "teaching assistant (TA) chatbots" for academic use. The study focuses on two major development paradigms: low-code platforms (e.g., AnythingLLM, Botpress) that enable rapid deployment with limited customization, and custom-coded solutions (e.g., using LangChain, FAISS) that offer greater flexibility, extensibility, and control [7]. Key components examined include LLM integration methods, memory architectures, RAG pipelines, vector-based document indexing, and infrastructure options for both cloud and self-hosted setups.

The primary objective is to establish a practical, comparative framework that helps developers and institutions select the most appropriate chatbot development approach based on factors such as scalability, data privacy, maintenance, and pedagogical alignment. Although this paper emphasizes applications in computing education, the findings are applicable to other knowledge-intensive domains. Notably, topics such as LLM pretraining, fine-tuning, or multimodal interface design are acknowledged but remain outside the current study's implementation-focused scope. This work is motivated by the need to align emerging LLM technologies with educational theory and classroom needs, ensuring that chatbots not only deliver technical functionality but also support meaningful learning experiences through well-informed pedagogical design.

## II. Background Work

### A. Literature Review

Recent studies have explored the integration of LLMs such as GPT-3.5, ChatGPT, and Google Bard into educational chatbots. Yigci (2023) [8] highlights their potential for asynchronous learning and personalized feedback, while also addressing issues like hallucination and bias. Schmucker et al. (2024) [6] presents a tutoring system that uses structured prompts for real-time student support, and Favero et al. (2024) [9] demonstrates how a Socratic chatbot can foster critical thinking. Yan et al. (2023) [10] provides a broader perspective, reviewing models like GPT, BERT, and T5 in tutoring and grading applications, emphasizing the importance of adaptable and explainable systems in education.

Retrieval mechanisms and data handling are discussed to a lesser extent but remain critical. SpringerOpen (2023) critiques proprietary platforms for limiting retrieval flexibility, suggesting tools like FAISS for improved document search. While technical specifics are often missing, studies such as Schmucker et al. (2024) imply backend retrieval capabilities that support context-aware feedback. NLP pipelines, though not directly detailed, are assumed in all systems, and concerns about transparency in data ingestion and preprocessing are raised by Yan et al. (2023) [10] and SpringerOpen (2023), advocating for more developer control in academic deployments.

Dialogue management is a consistent focus across the literature. Yigci (2023) emphasizes the role of Prompt engineering in guiding educational responses, while Carbonel and Jullien (2024) [11] examine how LLMs shift instructor-student dynamics, necessitating thoughtful interaction design. Structured dialogue strategies are evident in the tutoring frameworks of Schmucker et al. (2024) [6] and Favero et al. (2024) [9] , showcasing the pedagogical impact of prompt chaining and feedback loops. However, user interface and backend implementation details are rarely discussed, though deployments in classroom settings suggest the use of lightweight web interfaces, potentially built with tools like Flask or Gradio.

Infrastructure, API integration, testing, and security are generally underreported. SpringerOpen (2023) [12] and Yan et al. (2023) [10] call for open, secure systems that support private deployment and institutional control, noting that current platforms lack transparency. While Wu, R., & Yu, Z. (2024) [13] identifies gaps in evaluation methods, and several studies report real-world deployments, few describe structured logging, analytics, or testing frameworks. These findings highlight the need for future work that not only focuses on educational outcomes but also documents the technical foundations required to build scalable, secure, and context-aware LLM-based chatbots.

## III. Key Components of LLM Chatbots

At the heart of the chatbot lies the core model, which serves as the central processing unit. This module consists of three key components that collectively handle language comprehension, contextual reasoning, and response generation. The intelligence and naturalness of the chatbot heavily depend on these components.

### A. Core Model

The chatbot's core model acts as its central processing unit and this directly determines the chatbot's effectiveness and conversational quality.

1) *Large Language Model:*
   The engine that interprets user input and generates meaningful responses.
2) *Prompt engineering:*
   Carefully designed prompts guide the model's behavior, ensuring accurate and relevant outputs.[14]
3) *Backbone Model:*
   Transformer-based architectures like GPT-4, PaLM 2, or LLaMA power the chatbot's reasoning and response formulation.

However, a language model alone is not enough user input needs to be preprocessed before reaching the model to ensure clarity and structure.

### B. Input Handling

Before the chatbot can generate a response, it must first process the raw user input. The *input-handling* module consists of two components that ensure incoming text is properly structured and formatted for the Core Model.

1) *Text preprocessing:*
   Removes unnecessary characters, normalizes text, and handles spelling corrections. [15]
2) *Context Manager:*
   Maintains conversation history, tracks dialogue flow, and prevents redundant responses. [16]

Once the input is processed, maintaining continuity in conversation is essential. This is where the memory module comes into play.

### C. Memory

To make interactions feel natural, the chatbot must retain information across exchanges. The *Memory* module is composed of three components that enable both short-term and long-term recall, ensuring continuity in conversations.

1) *Short-Term Memory:*
   Temporarily stores recent exchanges to maintain conversational flow within a session.
2) *Long-Term Memory:*

Retains historical data, user preferences, and past interactions for a more personalized experience.

3) *Memory Retrieval Engine:*
Fetches relevant stored data to provide context-aware responses.

However, even with memory, the chatbot may not always have sufficient knowledge within its training data. To address this, it must be able to retrieve external information dynamically.

### D. Knowledge and Retrieval

A chatbot's effectiveness improves significantly when it can access real-time data and external knowledge. The *knowledge and retrieval* module enhances response accuracy by integrating both internal and external information sources. It comprises two major components.

1) *Knowledge Base:*
Internal sources like policy documents, product manuals, or structured databases.

2) *RAG Pipeline:*
Retrieves relevant data and integrates it with LLM-generated responses for improved accuracy [17].

Once the chatbot has generated a response using both internal and external knowledge, it must ensure that the output is appropriate and well-structured.

### E. Response Management

Generating a response is only part of the equation ensuring that it is well-formatted, safe, and aligned with the conversation is equally important. The *Response Management* module consists of two key components that refine and moderate outputs before presenting them to the user.

1) *Response Post-Processing:*
Enhances responses by refining grammar, adding citations, and ensuring coherence.

2) *Safety and Moderation:*
Applies filters for toxicity detection, bias mitigation, and compliance with ethical guidelines [18].

Beyond just managing responses, a chatbot should also tailor interactions based on individual users to create a more engaging experience.

### F. Personalization

To make interactions more dynamic and relevant, the *personalization* module adapts responses based on user profiles, preferences, and past interactions. This module consists of two components that ensure responses feel tailored to each user.

1) *Contextual Prompts:*
Customizes prompts based on user behavior, improving relevance.

2) *Adaptive Response Generator:*
Modifies tone and content to match user expectations [19].

Despite these optimizations, the chatbot must continue evolving. Gathering feedback and refining the model ensures continuous improvement.

### G. Feedback and Learning

No chatbot is perfect from the start. The *feedback and learning* module consists of a single but crucial component that enables iterative improvements by incorporating user feedback and refining the model over time.

1) *Model Update Pipeline:*
Collects feedback data for fine-tuning and incremental enhancements.

To ensure all these processes function smoothly, the chatbot requires an efficient communication layer to interact with users across different platforms.

### H. Communication and Interface

The final layer of the chatbot system is the Communication and Interface module, which facilitates user interactions through various channels and integrations. It consists of three essential components that enable smooth communication between the chatbot and external systems.

1) *User Interface:*
Provides access via web-based chat interfaces, mobile applications, or voice assistants.

2) *API Layer:*
Connects the chatbot with external services, CRMs, and enterprise systems.

3) *Multi-modal Input Support:*
Enables users to interact using text, voice, or even image-based queries.

## IV. APPROACHES TO BUILDING LLM BASED CHATBOTS

Our study categorizes chatbot development approaches into two broad types: no-code platforms and custom-coded solutions. No-code tools such as Botpress, Landbot.io, Rasa, and AnythingLLM enable rapid deployment using visual interfaces, making them suitable for FAQs, customer service, and small-scale use cases without requiring programming skills [20, 21]. However, they limit model customization, integration flexibility, and advanced use of LLMs. In contrast, custom-coded chatbots built using frameworks like Flask, Node.js, LangChain, and open-source LLMs (e.g., LLaMA, GPT, Claude) offer greater control over performance, privacy, and architecture, essential for enterprise applications and RAG systems. This approach, however, demands more technical expertise and infrastructure investment.

### A. Low-Code LLM Chatbot

1) *Definition and Characteristics:*
Low-code platforms provide a balance between no-code simplicity and fully custom-coded flexibility. These tools allow users to build LLM-driven chatbots using easy-to-use interfaces, ready-made building blocks, and smooth integrations. One of their main strengths is handling complex backend tasks like connecting to databases or managing servers behind the scenes, so users don't have to deal with them directly. They also simplify how the chatbot communicates with the language model, using tools like visual builders, which let users design chatbot conversations by dragging and connecting blocks (similar to flowcharts), and prompt management systems, which help organize and adjust the instructions sent to the language model.

*2) Need for a Framework :*

Many platforms also support RAG, allowing the chatbot to pull answers from uploaded documents or knowledge bases. While these platforms may limit deep customization, they strike a useful balance between control and ease of use, enabling quick testing and deployment of chatbots [20]. Many low-code platforms are available in the market today, but we have selected Botpress, AnythingLLM, Rasa, Landbot.ai, and Golem.ai for this study due to their popularity, community support, enterprise usage, and variety in design philosophy—ranging from open-source solutions to visual-first tools with commercial deployment support.

*3) Common Frameworks :*

*Botpress* is an open-source, modular platform tailored for building conversational agents via visual editors. It supports intent recognition, natural language understanding (NLU), and API-based integrations. Notably, it includes LLM-enhanced responses through GPT-based models and supports multi-channel deployment (e.g., Slack, Messenger). Developers can refine user flows using memory storage and flow logic [22].

*AnythingLLM* emphasizes quick setup of local or hosted LLM-based chatbots, especially for knowledge retrieval. Users can upload PDFs, documents, or URLs, which the platform indexes for RAG interactions. It features a configuration dashboard for vector stores and LLMs (e.g., OpenAI, Claude, LLaMA) and supports self-hosted options like Ollama, making it particularly appealing for privacy-sensitive deployments [23].

*Rasa*. Though more technical by design, Rasa's Rasa X and Pro versions introduce low-code capabilities via GUI dialogue editors, prebuilt connectors, and integrated data pipelines. It is open-source and built around NLU and dialogue management. Its modular structure supports gradual evolution from low-code use to advanced custom solutions, catering well to scalable enterprise needs [24].

*Landbot.ai* offers a highly visual, drag-and-drop interface targeting customer engagement and marketing. It supports conditional logic, OpenAI-based GPT responses, and basic API integrations. Although its backend extensibility is limited, Landbot is ideal for fast deployment of interactive, branded web and messaging experiences [25]

*Golem.ai* stands out with its emphasis on explainability and ethical AI. Unlike black-box LLM systems, it merges rule-based logic with AI-driven responses, offering transparency in conversational flows. This makes it suitable for compliance-heavy sectors. The platform also features multilingual support and integrations with enterprise tools like Customer Relationship Management (CRMs) and document systems [26]. Following Table 1. has comparison between different parameters of the above frameworks.

*4) Advantages and Limitations :*

Low-code platforms significantly reduce development time and lower the barrier to entry for teams without extensive programming expertise. Many come with built-in LLM integrations, analytics, and tools for rapid iteration, making them suitable for experimentation and early-stage deployments. However, limitations exist. Flexibility is constrained to predefined logic and supported models. Proprietary tools may introduce privacy concerns, and fine-grained control over Prompt engineering or performance tuning is limited. These drawbacks make such platforms less suitable for high-scale, highly customized, or data-sensitive applications.

To better understand their distinctions, a comparison table highlights deployment models (cloud or local), supported LLMs, file ingestion limits, customization levels, and target use cases.

TABLE 1. COMPARISON BETWEEN LOW-CODE PLATFORMS

| Platform | Rasa | Botpress | Golem.ai | AnythingLLM | Landbot.io |
|---|---|---|---|---|---|
| Type of Platform | Open-source chatbot framework | Low-code chatbot platform with automation | AI-powered decision-making chatbot | Open-source document-based chatbot | Drag-and-drop chatbot builder for marketing |
| Cloud or Local? | Both (Local & Cloud) | Both (Local & Cloud) | Cloud-only (No self-hosting) | Both (Local & Cloud) | Cloud-only (No local hosting support) |
| Max File Size Allowed (MB) | No direct support; requires external OCR or NLP preprocessing | 10 MB (PDFs require preprocessing) | 5 MB (Optimized for structured text) | 50 MB (Supports structured PDFs, text, and embedding models) | Not supported (Focused on structured chat flow, no file handling) |
| Backbone Model | Custom NLP models (SpaCy, TensorFlow, Hugging Face) | Supports GPT-4, Claude, Cohere, and other LLMs | Proprietary NLP model trained for structured decision-making | User-defined LLMs (Default: Ollama; Supports GPT-4, LLaMA, Falcon) | Proprietary NLP model optimized for customer service |
| Input Token Limits | Unlimited (Self-hosted); limited by cloud provider if using API | 8,192 tokens (GPT-4); other LLMs vary | 50,000 characters per document | 4,096 tokens (Ollama); up to 32,768 tokens (GPT-4 with API) | 2,048 tokens per interaction |
| Credit Limits | No limits for self-hosted; Cloud API usage varies by plan | Free tier: 2,000 messages/month; Higher tiers increase limits | Subscription-based; enterprise pricing includes usage credits | Self-hosted: No limits; Cloud: Paid tiers with increasing limits | Subscription-based; message limits vary per plan |
| Cost | Free (Self-hosted); Enterprise version available for custom pricing | Free to $495/month (teams); Enterprise pricing available | Custom pricing based on enterprise use cases | Free (self-hosted); Cloud plans start at $50/month | Starts at $30/mo.; Enterprise plans available |
| Requirements Other than Platform | Requires on-premise servers or cloud deployment (AWS, GCP, Azure) | Requires local deployment with Docker/Kubernetes or cloud hosting | Requires API integration with external services for automation | Requires server setup or cloud hosting (AWS, Azure) with vector database (Chroma DB, Pinecone) | Requires API integration with third-party services (Salesforce, HubSpot, WhatsApp) |
| Other Explanation | Best for developers who need full control over chatbot logic and data privacy | Ideal for teams needing multi-LLM integration with visual chatbot builder | Best suited for structured text processing and automation workflows | Designed for research-driven document chatbots with RAG capabilities | Best for non-technical users who need marketing chatbots with visual logic-based workflows |

## B. Custom-Coded LLM Chatbot

### 1) Definition and Characteristics

Custom-coded LLM chatbots are built using flexible, programmable frameworks that allow full control over system architecture and behavior. Unlike low-code platforms that abstract away the internal logic, custom-coded solutions enable developers to manage key components such as Prompt engineering, retrieval logic, context handling, and memory persistence [27, 28]. These systems support advanced features like multi-turn memory, Retrieval-augmented generation (RAG), and modular prompt pipelines that dynamically adapt to user context [29].

This level of customization is crucial for educational chatbots, where domain-specific logic such as grading policies or course rules must be incorporated into chatbot responses. Custom-coded solutions allow for integration with multiple LLMs, dynamic document parsing, fine-grained memory control, and improved handling of personalized or complex queries [28, 30]. While low-code platforms often limit retrieval to simple keyword matches or prebuilt APIs, custom-coded designs offer token-level control, enabling intelligent chunking, adaptive retrieval, and the use of vector databases for semantic search [29].

### 2) Need for a Framework

Building a robust LLM-based chatbot from scratch involves orchestrating multiple components such as memory, LLM APIs, document search, and session handling [27, 30]. To manage this complexity, developers rely on orchestration frameworks that streamline the integration of these elements into cohesive pipelines. Unlike low-code solutions, which offer fixed workflows and limited extensibility, custom frameworks provide modular chaining, custom logic injection, and multi-agent routing [28, 29].

Frameworks such as LangChain, Haystack, and LlamaIndex play a central role in this ecosystem. These tools simplify the construction of flexible pipelines that support conditional logic, custom retrievers, long-term memory buffers, and real-time document querying—all of which are essential for educational chatbots that need to manage evolving conversations over extended sessions [28, 30].

*3) Common Frameworks*

*a) Orchestration and Prompt Chaining*

- LangChain — A comprehensive framework designed to manage complex chatbot workflows. It supports prompt chaining, memory modules, and tool integration (e.g., APIs, databases). LangChain is well-suited for chatbots that require context-aware dialogue, retrieval from external sources, or multi-step reasoning [28].

- Haystack — A powerful toolkit focused on building retriever–reader pipelines for accurate response generation. It is ideal for scenarios where chatbots need to extract relevant information from long documents or knowledge bases using RAG [30].

*b) Data Ingestion and Knowledge Indexing*

- LlamaIndex — Enables the ingestion of structured and unstructured data (e.g., PDFs, notes) into a searchable knowledge base. LlamaIndex is optimized for use cases where large collections of course content need to be accessed efficiently [31].

*c) Frontend and Interface Layer*

- Gradio / Fast API — For user-facing interaction, Gradio provides a fast and easy way to prototype chatbot interfaces using Python, making it ideal for small-scale or demo applications [32]. For more scalable deployments, FastAPI (or Flask) supports features such as user authentication, logging, and API-based communication, which are essential for institution-level educational bots.

The above frameworks are compared between different parameters in Table 2.

*4) Advantages and Limitations*

Custom-coded LLM chatbots offer significant advantages in flexibility, precision, and scalability. Developers can implement smart document chunking, adaptive retrieval mechanisms, long-term vector-based memory, and multi-model orchestration [29, 31]. These features enable chatbots to deliver highly accurate, context-sensitive responses, tailored to institutional policies and academic content. Furthermore, custom implementations can ensure data privacy and security through on-premise hosting or encrypted data flows essential for compliance with regulations like FERPA [33].

However, these benefits come with increased complexity. Custom solutions require proficiency in Python, API design, containerization (e.g., Docker), and LLM orchestration. Maintenance also poses challenges, including dependency management, infrastructure scaling, and performance optimization. Additionally, mastering concepts such as prompt templating, vector search, and session memory can introduce a steep learning curve for development teams [30].

To highlight the strengths of custom-coded platforms, the following table summarizes key features such as flexibility, memory control, token customization, and suitability for advanced educational use.

TABLE 2. COMPARISON BETWEEN CUSTOM-CODE PLATFORMS

| Platform | LangChain | LlamaIndex | Haystack | Gradio + FastAPI/Flask |
|---|---|---|---|---|
| Type of Platform | Orchestration Framework | Data Indexing & Retrieval Framework | Retriever–Reader QA Pipeline Framework | Frontend UI + Backend API Layer |
| Cloud or Local? | Both (Cloud-compatible, often local for development) | Both (Depends on backend storage and vector DB) | Both (Cloud deployment supported, local dev common) | Both (local prototyping and cloud deployment supported) |
| Max File Size Supported | Depends on vector DB and memory configuration | Up to hundreds of MB (based on indexing method) | Varies by retriever and storage backend | N/A (depends on underlying backend and API design) |
| Backbone Model | Supports OpenAI, Cohere, Claude, LLaMA, others via API | Model-agnostic, used in conjunction with LangChain or similar | Compatible with Hugging Face, OpenAI, Cohere, LLaMA, etc. | Model-agnostic; integrates with any LLM through APIs |
| Token Limits | Depends on integrated LLM (e.g., 4k–128k tokens) | Inherited from LLM used (via LangChain or direct) | Depends on LLM provider and retriever configuration | Depends on the connected LLM (typically 4k–128k tokens) |
| Credit Limits | Subject to usage of API-based LLMs | N/A unless connected to paid LLM APIs | Usage-bounded if external APIs are involved | Bounded by the usage policies of the LLM or hosting platform |
| Cost | Open-source; API usage may incur charges | Open-source; vector DB or LLM APIs may incur cost | Open-source; cost depends on APIs and infra | Free (open-source); hosting or API usage may incur costs |

| Requirements | Python, API keys for LLMs, vector DB integration (e.g. FAISS) | Python, document parsers, vector DB backends | Python, retrievers/readers, model APIs | Python, ASGI server, UI framework, backend logic, optional Docker |
|---|---|---|---|---|

## V. COMPARATIVE ANALYSIS: LOW-CODE VS. CUSTOM CODED CHATBOTS

The decision between using a low-code platform and pursuing a custom-coded approach is central to chatbot development, especially when building LLM-powered systems for specific domains like education. Each method comes with its own trade-offs across development time, customization potential, cost, and scalability.

*Development Time and Ease of Implementation*

Low-code platforms such as Botpress, Landbot.ai, and AnythingLLM significantly reduce the initial development time by offering pre-built modules, graphical user interfaces (GUIs), and native LLM integrations [22, 23]. These platforms are designed to help non-developers quickly deploy conversational agents with minimal configuration. For instance, Landbot's drag-and-drop interface allows educators to launch simple course assistants without writing a single line of code, Likewise, AnythingLLM offers plug-and-play document ingestion, vector storage, and OpenAI integration, making it ideal for rapid prototyping without heavy infrastructure setup.

In contrast, custom-coded implementations require a deeper technical foundation. Developers must manage LLM integration, memory architecture, vector-database configuration (e.g., FAISS), and frontend/backend orchestration using tools like LangChain, FastAPI, or Flask. Although this approach demands more time and expertise initially, it enables the creation of highly tailored systems suited to an institution's specific pedagogical workflows.

*Customization and Flexibility*

Customization is where low-code platforms reach their limitations. While platforms like Botpress and Golem.ai allow conditional logic and limited scripting, the developer is constrained by the platform's built-in capabilities and available APIs. For example, while Golem.ai supports natural language rule customization, it may not support hosting self-trained LLMs or custom memory strategies.

Custom-coded systems, however, allow full control over how chatbots interpret, retrieve, and generate responses. Developers can fine-tune LLMs on institution-specific data, design custom RAG pipelines, and build secure authentication layers. This level of control is essential when integrating proprietary data sources, supporting large document inputs, or optimizing latency for real-time interaction.

*Cost Considerations (Short-Term vs. Long-Term)*

Low-code platforms typically operate on subscription-based pricing models, offering free tiers with limited usage. In the short term, they are cost-effective, especially for pilot projects or lightweight deployments. For instance, Rasa's open-source core is free to use, but advanced features and enterprise support are locked behind commercial plans. Platforms like AnythingLLM that support local hosting provide hybrid cost flexibility—avoiding monthly LLM usage fees while still abstracting away infrastructure complexity.

Custom-coded solutions may incur higher upfront costs, especially when deploying self-hosted LLMs or provisioning cloud infrastructure. However, in the long run, they can be more economical for institutions with specific needs, high usage volumes, or privacy restrictions that require local deployments. Furthermore, open-source tools offer cost-saving opportunities if managed efficiently.

*Performance and Scalability Trade-Offs*

Scalability is another key differentiator. Low-code platforms offer limited performance tuning and may impose restrictions on file size, number of concurrent users, or supported model types. As demand scales—e.g., hundreds of students interacting simultaneously—platform limitations can become bottlenecks.

Custom implementations are more scalable when designed correctly. Developers can optimize API calls, implement caching strategies, and even switch LLM providers based on response quality or pricing. They can also deploy models locally for faster inference and tighter data control. This flexibility is essential in educational environments where latency and uptime directly impact the learning experience.

To guide the selection of an appropriate chatbot development strategy, our comparative framework recommends evaluating each approach across four primary dimensions identified through our research: technical expertise availability, required customization depth, deployment scale, and data privacy concerns. Specifically, institutions with limited technical resources, smaller-scale deployments, or initial experimentation needs would benefit from low-code platforms due to ease-of-use and faster prototyping. Conversely, institutions prioritizing pedagogical customization, scalability for large cohorts, advanced retrieval mechanisms (RAG), and strict data privacy would be best served by custom-coded solutions. Using these clearly defined dimensions, developers and educators can systematically assess institutional priorities against each chatbot development approach to make informed, strategic decisions.

## VI. CHALLENGES IN DEVELOPING LLM-BASED CHATBOTS

Building effective LLM-based chatbots requires more than just connecting to models like GPT-4 or DeepSeek. Developers must design systems that manage real-time conversations, retrieve accurate information from large content sources, and support memory across sessions [34]. Frameworks such as LangChain help orchestrate prompt logic and memory [35], while LlamaIndex and Haystack handle document retrieval and knowledge indexing. These tools must work together smoothly to ensure the chatbot delivers useful and context-aware responses.

Privacy and infrastructure are also major challenges. Using OpenAI's API or other cloud-based LLMs offers convenience but raises concerns about data security in educational settings. Alternatively, self-hosted deployments using models like LLaMA or DeepSeek require significant compute resources and setup [36]. Managing these trade-offs—especially when balancing speed, cost, and data protection—is crucial when developing chatbots for academic environments.

## VII. Future Work

Our research provides critical insights into the strengths and limitations of current chatbot development approaches, emphasizing the need for deeper pedagogical alignment, adaptive learning support, and enhanced interactivity. Building directly on these findings, our future work will focus on two impactful directions. First, integrating multimodal input processing—such as images, code snippets, and diagrams—within our chatbot pipelines, using frameworks like LangChain. This advancement is particularly crucial for improving instructional clarity and effectiveness in STEM disciplines, where visual representations greatly enhance cognitive engagement.

Second, we will prioritize implementing robust personalization and adaptivity features by deploying long-term learner models and real-time feedback loops via platforms such as FastAPI and Flask. By further leveraging techniques like Reinforcement Learning from Human Feedback (RLHF) and adaptive retrieval strategies with LlamaIndex, we aim to create truly responsive, individualized tutoring experiences. Collectively, these enhancements will significantly advance our chatbot's educational impact, transforming it from a primarily informational tool into a personalized learning partner that deeply aligns with diverse learners' needs, preferences, and cognitive processes.

## VIII. Conclusion

This research has examined the development of course-specific chatbots utilizing Large Language Models (LLMs), comparing low-code platforms like AnythingLLM and Botpress with custom-built solutions powered by frameworks such as LangChain, Haystack, and LlamaIndex. While low-code tools facilitate rapid deployment, they often constrain customization in areas like memory management, prompt flexibility, and retrieval logic, rendering them less suitable for complex educational applications [40].

In contrast, custom-coded solutions empower developers to integrate tools like ChromaDB, Gradio, FastAPI, and Flask, providing comprehensive control over the user interface, secure data handling, and backend model orchestration. These systems support advanced features such as Retrieval-augmented generation (RAG), dynamic prompts, and scalable deployment, making them ideal for educational settings that demand accuracy, adaptability, and privacy [41].

As these technologies evolve, hybrid approaches are likely to become the standard, combining the ease of use of low-code platforms with the powerful customization capabilities of custom solutions to support next-generation, learner-centered education [42].